\documentclass{ptapap}

\usepackage{amssymb}

\author{Petra Sukov\'a}[CFT]
\author{Szymon Charzy\'nski}[WU,CFT]
\author{Agnieszka Janiuk}[CFT]
\affil[CFT]{Center for Theoretical Physics PAS\\
  Al. Lotnikow 32/46, 02--668 Warszawa, Poland}
\affil[WU]{Chair of Mathematical Methods in Physics, University of Warsaw\\ ul. Pasteura 5, 02-093 Warszawa, Poland}

\title{Relativistic low angular momentum accretion in 3D}

\begin{document}

\maketitle

\begin{abstract}

We study the low angular momentum accretion of matter onto Schwarzschild or Kerr black hole using fully relativistic numerical simulations. Our aim is to include the rotation of both the gas and the black hole and the presence of the magnetic field into the picture. We compare resulting structure of transonic flow with results of 1D pseudo-Newtonian semi-analytical computations of non-magnetized flow.

\end{abstract}

\section{Introduction}

The problem of low angular momentum accretion on the black hole does not have an analytic solution, analogous to the fully relativistic Bondi solution for spherically symmetric accretion. Therefore it has to be treated with numerical methods.
In our simulations, we focus on the interplay between adopted parameters, like the specific angular momentum $\lambda$ or adiabatic index $\gamma$ and the evolution of the flow.
We also investigate the possible existence of standing shocks in such flows. Our computations are relevant in the case, when the sub-Keplerian component is important for explaining the properties of the accretion flow. 
One of the physical set-up suitable for our computations
are the microquasars, in which two component advective model together with Propagating Oscillatory Shock model was used for describing the evolution of the source and the frequency of its quasi-periodic oscillations during outburst. 
Another physical situation is the accretion onto supermassive black holes in the low luminous galactic nuclei (including Sgr A*) in which the stellar winds can collide and loose most of its angular momentum far away from the center.
In this case, the observations made at milimeter wavelength, show the presence of strong magnetic fields threading the accreting gas in our Galaxy center.
Our future simulations are planned to add also the magnetic field into the picture.

\section{Low dimensional computations}

We have carried out simulations of one-dimensional hydrodynamical model of shocked quasi-spherical accretion flow with a constant angular momentum (hereafter 1D model), which confirm the shape of the semi-analytical steady solution and its dependence on the leading parameters (angular momentum $\lambda$, energy $\epsilon$ and adiabatic index $\gamma$).
The simulations yield the shock front unstable for a subset of parameters which led to oscillations of the shock front around the position given by the steady solution. The frequency depended on the distance from the center
(\cite{Sukova21022015}, hereafter [SJ]),
hence on the value of angular momentum.

\section{Three dimensional numerical simulations}

In order to verify the predictions of simplified 1D model we have performed higher dimensional computations in full relativistic and 3D framework.
We used the \texttt{Einstein Toolkit} computational package\footnote{\tt http://www.einsteintoolkit.org} \citep{2012CQGra..29k5001L}.
The background spacetime is stationary Kerr solution parameterized by Kerr-Shield coordinates. Fixed mesh with 8 refinement levels (grid spacing from $12.8M$ to $0.1M$) is used in order to cover required amount of space and at the same time to resolve the region near the event horizon with sufficient accuracy. The computation domain is a box with coordinates ranging from $-1024M$ to $1024M$ in all three dimensions and the black hole with $a=0$ is located in the center at $[0,0,0]$.

The initial conditions of the hydrodynamical variables (including $u^r$) are set according to analytical Bondi solution \citep{2014CQGra..31a5005M}, parameterized by the position of the sonic point $r_{\rm s}$ and the mass accretion rate $\dot{M}$. These conditions are modified by adding a nonzero $\phi$ component to the four-velocity of the gas in the outer region, which is computed according to
\begin{eqnarray}
u^\phi = \frac{\lambda}{r^2} \sin^2{\theta}, \qquad r>r_{\rm b}, \label{uphi}\\
u^\phi = 0 ,\qquad r<r_{\rm a},
\end{eqnarray}
in the Boyer-Lindquist coordinates. Between $r_{\rm a}$ and $r_{\rm b}$ the values are smoothed by a cubic spline. The time component of four-velocity is set from the normalization condition assuming $u^\theta=0$.
Such prescription is used in order to mimic the ``quasi-spherical'' condition used in [SJ] for the flow with constant $\lambda$. However, in 3D we need to deal with the problem of this assumption near the axis of rotation. The vanishing of angular momentum at the axis of rotation is provided by the $\sin^2\theta$ factor in (\ref{uphi}).

The evolution of the gas is simulated by the GRHydro module \citep{2014CQGra..31a5005M} supplied with our own modifications. It is equipped with the shock-capturing routines and it solves the Euler equation for a perfect fluid in the ADM 3+1 formulation. 

We follow the behavior of the gas inside the box while we keep the values in the boundary zones constant during the evolution.
We compare with the 1D model the radial Mach number $\mathfrak{M}=w^{\rm rad}/c_{\rm s}$,
where $c_{\rm s}=\sqrt{\frac{\gamma p}{\rho(1+\epsilon)+p }}$ is the local sound speed relative to the fluid,
$\rho$ is the fluid rest mass density, $\epsilon$ is specific internal energy and $p$ stands for the gas pressure. Moreover,
$w^{\rm rad}$ is the radial velocity computed from the Cartesian components $w^i=(\alpha v^i - \beta^i) W/\alpha$ with $\alpha$, $\beta^i$ being the lapse and shift of the Kerr-Shield coordinates,
$W=(1-v^i v_i)^{-1/2}$ is the Lorentz factor of the flow and $v^i$ is the fluid three-velocity as seen by the Eulerian observer.

\begin{figure}
  \centering
    \includegraphics[width=0.514\textwidth]{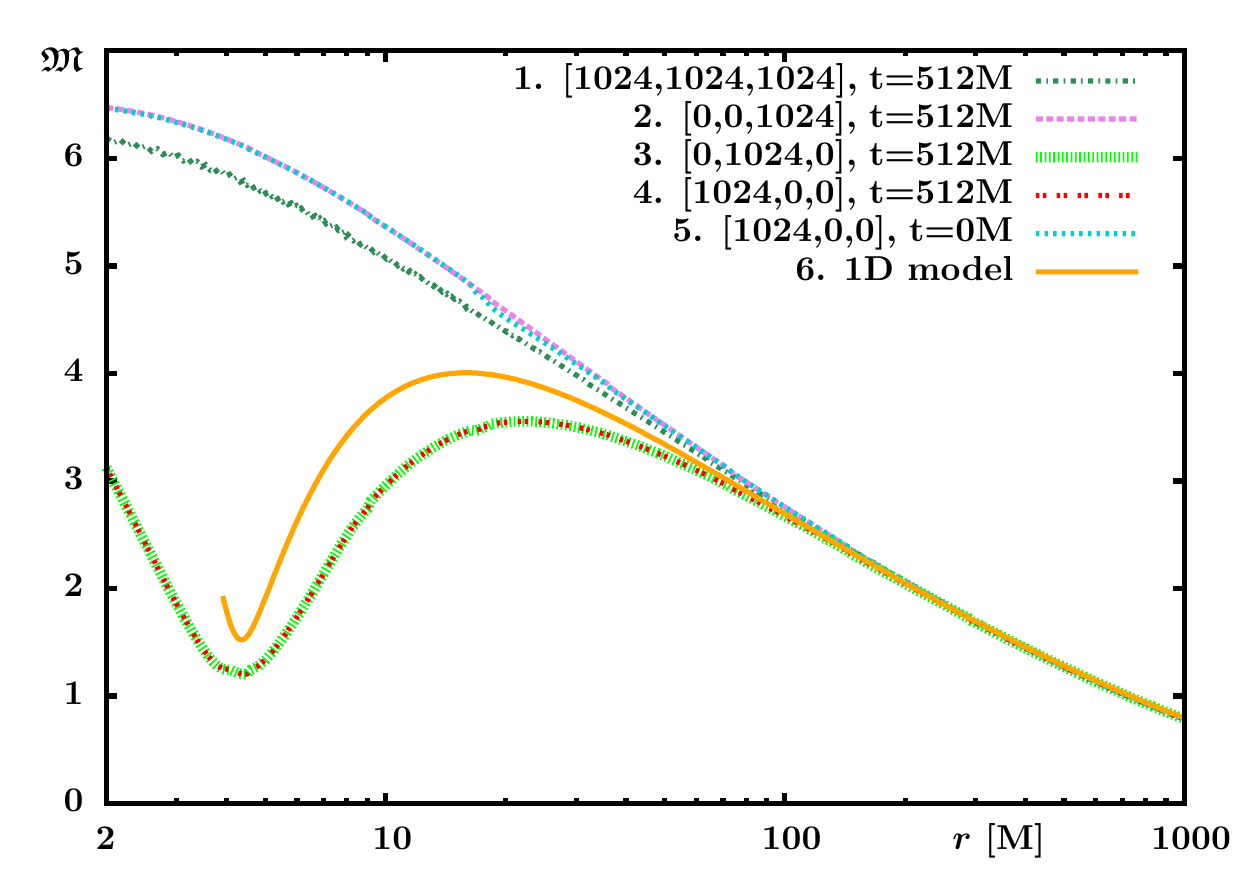}
    \includegraphics[width=0.466\textwidth]{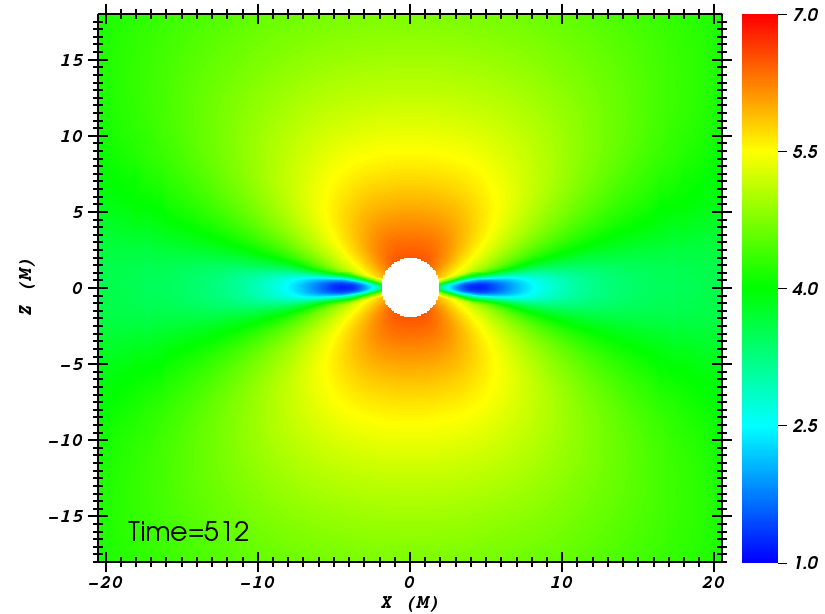}
    \caption{\label{Sukova_fig_1} Left: profile of radial Mach number $\mathfrak{M}$ along the lines starting at [0,0,0] and pointing along axes (x-axis -- line 4, y-axis -- line 3, z-axis -- line 2) and the diagonal (line 1). Line 5 shows the initial state and line 6 is the output of the 1D model. Right: xz-slice showing $\mathfrak{M}$ in the innermost region.}
\end{figure}

The exemplary result is presented in Fig.~\ref{Sukova_fig_1}. In this simulation we set $r_{\rm s}=732.3415M$, $r_{\rm a}=15M$, $r_{\rm b}=20M$ and $\lambda=3.79M$. These parameters correspond to our previous 1D model with the outer critical point given by $\lambda=3.79M, \epsilon=0.001, \gamma=4/3$.
In Fig.~\ref{Sukova_fig_1} we show the profile of $\mathfrak{M}$ along several lines starting from the point [0,0,0] and crossing the grid in different directions (e.g. along the three axes, as labeled by their end points).  At the initial time, the profile of $\mathfrak{M}$ corresponds to the spherically symmetric Bondi solution, so the profiles in all directions coincide.
After the start of the simulation, the rotating gas reaches the inner region close to the black hole and the profile of the Mach number close to the equatorial plane resembles the one from the 1D model, which crosses the outer critical point.

\section{Discussion}
In the simulation with initial conditions given by the Bondi solution perturbed by small rotation (line 5), the accreting gas settles to the solution going through the outer critical point and qualitatively resembles the 1D model given in [SJ] (line 6).
Also this resemblance is valid only close to the equatorial plane (lines 3, 4) and the profile is more like Bondi solution near the rotation axis (lines 1, 2). The other variables ($\rho$, $p$) are no longer spherically symmetric, so that the ``quasi-spherical'' assumption from [SJ] is violated. Absence of spherical symmetry together with full GR framework in 3D simulation is a source of quantitative differences between 3D model and 1D pseudo-Newtonian approach presented in [SJ].

In order to observe the shock in the flow 
the initial condition in the inner region has to be set close to the inner branch of solution (as in [SJ]). Work on such a simulation is in progress. Next step is to add the spin of BH and magnetic field into the picture.
Magnetic fields, due to the MHD turbulence, will provide an internal mechanism for the angular momentum transport within the accreting flow.
Otherwise, the conditions for the time-dependent shock phenomena are still too idealistic given numerical evidence from the past work. Only the full three-dimensional treatment of magnetic fields is physically robust.

\acknowledgements{We thank Oleg Korobkin for helpful discussions. We acknowledge support from Interdisciplinary Center for Computational Modeling of the Warsaw University and National Science Center (2012/05/E/ST9/03914).}

\bibliographystyle{ptapap}
\bibliography{Sukova}

\end{document}